\documentclass[aps,pra,twocolumn,superscriptaddress,longbibliography]{revtex4-1}

\usepackage{latexsym}		
\usepackage{amsmath}
\usepackage{graphics}
\usepackage[usenames]{color}
\usepackage{color}
\usepackage{tikz}
\usepackage{amssymb}

\usepackage{braket}

\begin{document}
\newcommand{\pd}[2]{\frac{\partial #1}{\partial #2}}
\newcommand{\der}[2]{\frac{d #1}{d #2}}
\newcommand{\pdd}[2]{\frac{\partial^2 #1}{\partial #2^2}}
\newcommand{\secder}[2]{\frac{d^2 #1}{d #2^2}}

\title{Complex collisions of ultracold molecules: a toy model}
\author{Jia K. Yao} \email{jyyao@caltech.edu}
\affiliation{Department of Physics and Astronomy, Rice University, Houston, Texas 77005, USA}
\affiliation{Rice Center for Quantum Materials, Rice University, Houston, Texas 77005, USA}
\author{Nirav P. Mehta}
\email{nmehta@trinity.edu}
\affiliation{Trinity University, One Trinity Place, San Antonio, TX. 78212-7200 USA}
\author{Kaden R.~A. Hazzard} \email{kaden.hazzard@gmail.com}
\affiliation{Department of Physics and Astronomy, Rice University, Houston, Texas 77005, USA}
\affiliation{Rice Center for Quantum Materials, Rice University, Houston, Texas 77005, USA}
\date{\today}

\begin{abstract}
We introduce a model to study the collisions of two ultracold diatomic molecules in one dimension interacting via pairwise potentials. We present results for this system, and argue that it offers lessons for real molecular collisions in three dimensions. We analyze the distribution of the adiabatic potentials in the hyperspherical coordinate representation as well as the distribution of the four-body bound states in the adiabatic approximation (i.e. no coupling between adiabatic channels). It is found that while the adiabatic potential distribution transitions from chaotic to non-chaotic as the two molecules are separated, the four-body bound states show no visible chaos in the distribution of nearest-neighbor energy level spacing. We also study the effects of molecular properties, such as interaction strength, interaction range, and atomic mass, on the resonance density and degree of chaos in the adiabatic potentials. We numerically find that the dependence of the four-body bound state density on these parameters is captured by simple scaling laws, in agreement with previous analytic arguments, even though these arguments relied on uncontrolled approximations. This agreement suggests that similar scaling laws may also govern real molecular collisions in three dimensions. 
\end{abstract}

\maketitle

\section{introduction}
Ultracold molecules have emerged as a new platform for quantum science and technology~\cite{Koch:review_control,Gadway_2016}. They combine the unique, tunable, coherent setting of ultracold matter with strong dipolar interactions and numerous stable internal rotational and vibrational states. Consequently, this platform has wide-ranging applications, including exploring new phases of matter and nonequilibrium behavior~\cite{Lepers:Dipole, wall:quantum_2015, hazzard:many-body_2014, yan:realizing_2013, carr:cold_2009, baranov:condensed_2012, bohn:cold_2017, lemeshko:manipulation_2013,Blackmore2018}, enabling quantum computation~\cite{demille:quantum_2002, andre:coherent_2006, yelin:schemes_2006, herrera:infrared_2014, karra:prospects_2016,ni:dipolar_2018}, performing precision measurements, such as measuring the electron electric dipole moment~\cite{kozlov:parity_1995, flambaum:enhanced_2007, hudson:improved_2011, baron:order_2014, cairncross:precision_2017}, and studying chemical reactions in the quantum regime~\cite{ospelkaus:quantum_2010, ni:dipolar_2010,miranda:controlling_2011, balakrishnan:perspective_2016, krems:molecules:2005, krems:cold_2008, liu:building_2018, guo:collision_NaRb, Meyer:bialkali_reaction}.

To fully realize these applications, it is necessary to theoretically understand the collisional behavior of molecules. For example, the density of bi-molecular bound states at the collision energy is important in determining the lifetime of a cloud of molecules~\cite{quemener:vibrational_2008,miranda:controlling_2011,ospelkaus:quantum_2010, Morita:cooling}, the many-body physics in an optical lattice~\cite{Docaj:2016prl,Wall:2017pra, Wall:2017pra2,ewart:bosonic_2018}, and chemical reaction rates. In addition to being necessary for applications, understanding molecular collisions is of fundamental scientific interest. An important question is whether and how quantum chaos occurs in molecular collisions: molecules are intermediate between atoms, which are mostly simple and non-chaotic, and thermodynamically large systems, which are frequently chaotic. It is believed that quantum systems which become chaotic in the classical limit display spectral fluctuations identical to those of random matrices generated from certain classes of statistical ensembles in random matrix theory~\cite{Berry:level_clustering,Bohigas:1984prl,MehtaML:2004random}. A chaotic time-reversal symmetric system is expected to be described by a Gaussian orthogonal ensemble (GOE), with a characteristic level repulsion described by the Wigner-Dyson distribution. On the other hand, levels of non-chaotic, integrable systems follow a Poisson distribution. Manifestations of random matrix theory are well-studied in nuclear spectra and collisions~\cite{Weidenmuller:2009rmp,Mitchell:2010rmp,Brody:1981rmp}.

Molecular collisions are much more complex than atomic collisions, because molecules have a much denser collection of electronic, rotational, vibrational, and hyperfine states, resulting in a several orders of magnitude higher density of resonances. This was suggested by Mayle \textit{et al.}~\cite{Mayle:2012pra,Mayle:2013pra}, and recently estimated by Christianen \textit{et al.}~\cite{Groenenboom:2019} to be $0.124\mathrm{\mu K^{-1}}$ for NaK + NaK collision complexes, without including the hyperfine states that will further increase the resonance density. Thus, there will exist many accessible closed-channel bound states around the collision energy -- even at the coldest available experimental temperatures. Consequently, while scattering resonances are routinely measured in ultracold atoms~\cite{Chin:RMP82.2015}, even for atoms with the densest resonance spectrum, such as that explored in lanthanide atoms (Dy and Er)~\cite{Frisch2014,Maier:2015pra, maier:emergence_2015,jachymski:analytical_2013,Jachymski:2015pra,jachymski:impact_2016,makrides:fractal_2018, augustovic:manifestation_2018, yang:classical_2017} and predicted in alkaline-earth-like atoms (Yb)~\cite{green:quantum_2016}, it is much harder to resolve the resonances in molecular collisions due to the significantly higher density of states. In addition to being difficult to experimentally measure, molecule-molecule collisions are extremely challenging to model, and quantitative calculations remain impossible for diatomic molecules. Recent work has emphasized the high resonance density inherent to bialkali (and heavier) atom-molecule~\cite{yang:observation_2018,croft:universality_2017, Croft:2017pra, frye:approach_2016,Docaj:2016prl,Wall:2017pra,Wall:2017pra2,Mehta} and molecule-molecule~\cite{ dawid:two_2018,Gregory2019:sticky,Mehta:boson,Mehta:few-body1D} collisions, building on earlier research on lighter molecules formed from lighter atoms~\cite{Forrey:1998pra, avdeenkov:ultracold_2001, bohn:rotational_2002, tscherbul:controlling_2006, simoni:ultracold_2006, quemener:vibrational_2008, tscherbul:magnetic_2009, simoni:feshbach_2009, zuchowski:reactions_2010, vonStecher:2009}. 

As a result of the experimental and theoretical challenges, open questions remain, such as: Is chaos universal in molecule collisions? What is the effect of the molecular size, the mass of the constituent atoms, and of the type of interactions?

In this paper, we present a simplified model of the collision of two identical molecules, which can shed light on the questions above while at the same time being amenable to straightforward numerical calculation. The principle simplification is to restrict the atoms to move in one dimension, interacting via a simple model potential (either Morse or P{\"o}schl-Teller). This potential is chosen to be deep enough to harbor many two-atom bound states, mimicking the situation in real molecules. For concreteness, we choose each molecule to be composed of two distinguishable fermionic isotopes. 

We find several results that may have implications for scattering of real molecules in three dimensions, which can be divided into three categories. First, we find that the statistics of the adiabatic potentials transition from chaotic to non-chaotic as the intermolecular separation increases, similar to results observed for atom-molecule scattering in calculations using realistic potential energy surfaces~\cite{frye:approach_2016}. We characterize how the detailed behavior of this crossover depends on the parameters of the model potential. We show that there are clear trends in the average spacing of the adiabatic potentials and the chaoticity of their level spacing statistics as a function of hyperradius, collision energy, molecular mass, and range of the potential. We expect that these qualitative trends persist to three-dimensional systems. 

The two remaining findings are more surprising. The second finding results from calculating the four-body bound states in the approximation that the adiabatic potentials are uncoupled. We find that the four-body level statistics show no apparent chaos even when the adiabatic potentials are strongly chaotic. This suggests caution when interpreting the level statistics of adiabatic potentials, which may not lead to corresponding statistics in the spectrum of the four-body complex, particularly if nonadiabatic couplings are weak or negligible. Lastly, and perhaps most interestingly, we derive a simple analytical relation between the four-body density of states and the two-atom interaction parameters. We show that, although the derivation of these analytic relations relies on uncontrolled approximations, they accurately capture the scaling with interaction parameters. This suggests that the analogous expressions for the three-dimensional system will allow one to simply estimate the effects of changing molecular species or other experimental parameters on complex molecule-molecule collisions.

The structure of this paper is as follows. Sec.~\ref{sec:II} describes the problem we solve and the numerical method, which uses the tools associated with hyperspherical coordinates. Sec.~\ref{sec:III} presents the results for the adiabatic potentials. Sec.~\ref{sec:IV} shows the density of four-body bound states in the adiabatic (zero-coupling) approximation from numerically solving the Schr{\"o}dinger equation, as well as their nearest-neighbor level statistics. It derives approximate analytic expressions for the scaling of the four-body density of states and shows that despite the uncontrolled approximations involved, they agree well with the numerical results. Sec.~\ref{sec:conclusions} summarizes the results, and gives an outlook, especially focusing on these results' relevance to real three-dimensional molecule-molecule scattering.

\section{Molecule-molecule scattering and hyperspherical coordinates}\label{sec:II}
We study a system of two identical diatomic molecules in one dimension. Each molecule consists of two fermionic atoms, giving two sets of identical fermions in total, labeled as (1, 2) and (3, 4). We take 1 and 3 to be distinguishable from 2 and 4. The masses of the distinguishable atoms are either set to be equal or different by a small amount. Throughout the paper, we fix the mass of atom 1 (and 2), i.e., $m_1 = m_2 = 1$, and vary the mass of atom 3 (and 4) to atom 1 by setting $m_3/ m_{1}= 1$ or $1.3$.

We present results for pairwise atom-atom interactions with two potentials: Morse and P\"{o}schl-Teller potentials. In either case, the model potential is assumed
to be identical for all pairs of atoms.
\begin{subequations}\label{eq1}
  \begin{gather}
U_{\mathrm{Morse}}(r) = D[1-e^{-a(r-r_{0})}]^2-D\label{1second}\\
U_{\mathrm{P\ddot{o}schl-Teller}}(r) = -\frac{D}{{\cosh(r/r_{0})}^2}\label{1third}
  \end{gather}
\end{subequations}
where $r \ge 0$ is the interatomic distance, $D$ is the depth of the potential, $r_0$ is the width of the potential, and $a$ is set to $1/r_0$, so that $U(0)/D$ is independent of $r_{0}$.

To numerically solve the Schr{\"o}dinger equation and interpret its eigenstates, we transform it into hyperspherical coordinates. Hyperspherical coordinates have been used widely~\cite{DELVES1958391,DELVES1960275,Macek:adiabatic, Bondi:Cl_MuCl, Chuluunbaatar:KANTBP, Lin:Hyperspherical_3body, Suno:three-body, Nielson:3body, Esry:hyperspherical,Mehta:N-body, Rittenhouse:four-fermion, croft:universality_2017} to study few-body physics and have proven to provide an efficient and powerful method for studying the full scattering and bound state problem. In this representation, the particle coordinates are rewritten into one hyperradial coordinate and a collection of angular coordinates (defined below).

In one dimension, the position and the mass of particle $i$ in the lab frame are labeled by $r_i$ and $m_i$, respectively. By setting the center-of-mass coordinate to zero, the remaining three degrees of freedom can be expressed in terms of Jacobi coordinates $\rho_1$, $\rho_2$, and $\rho_3$, defined as

\begin{subequations}\label{eq2}
  \begin{gather}
\rho_{1} = \sqrt{\frac{\mu_{12}}{\mu}}(r_{1}-r_{2})\label{2second}\\
\rho_{2} = \sqrt{\frac{\mu_{34}}{\mu}}(r_{3}-r_{4})\label{2third}\\
\rho_{3} = \sqrt{\frac{\mu_{12,34}}{\mu}}
\left(\frac{m_{1}r_{1}+m_{2}r_{2}}{m_{1}+m_{2}}-\frac{m_{3}r_{3}+m_{4}r_{4}}{m_{3}+m_{4}}\right),\label{2forth}
  \end{gather}
\end{subequations}
where 
\begin{subequations}\label{eq3}
  \begin{gather}
\mu_{12}=\frac{m_{1}m_{2}}{m_{1}+m_{2}}\label{3second}\\
\mu_{34}=\frac{m_{3}m_{4}}{m_{3}+m_{4}}\label{3third}\\
\mu_{12,34}=\frac{(m_{1}+m_{2})(m_{3}+m_{4})}{m_{1}+m_{2}+m_{3}+m_{4}}.\label{3forth}
  \end{gather}
\end{subequations}

\begin{figure}
\centering
\includegraphics[width=0.35\textwidth]{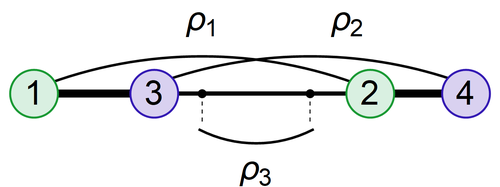}
\caption{\label{fig1} (color online) Jacobi coordinates $\rho_1$, $\rho_2$, and $\rho_3$ for the four-fermion system in one dimension, where the center-of-mass coordinate is set to 0. Particles 1 and 2 are indistinguishable, and particles 3 and 4 are indistinguishable. When (1,3) and (2,4) are coincident, and the separation between these molecules is large, the hyperspherical coordinates approach $(\phi,\theta) \rightarrow (\pi/4,\pi/2)$ for $m_1=m_2=m_3=m_4$.}
\end{figure}
Figure~\ref{fig1} illustrates these Jacobi coordinates for the system studied in this paper. The hyperspherical coordinates $(R,\phi,\theta)$ are related to the Jacobi coordinates by 
\begin{align}\label{eq4}
    \rho_{1} &= R \cos\phi \sin\theta\\\nonumber
    \rho_{2} &= R \sin\phi \sin\theta\\
    \rho_{3} &= R \cos\theta,\nonumber
\end{align}where the hyperradius is defined as
\begin{equation}\label{eq5}
    R = \sqrt{\rho_{1}^2+\rho_{2}^2+\rho_{3}^2}.
\end{equation}
In this formalism, the hyperradius $R$ effectively characterizes the size of the system, and the angular coordinates $\theta$ and $\phi$ describe the relative distances between pairs of atoms. For example, as $(\phi, \theta) \rightarrow (\pi/4,\pi/2)$, the system is separated into two constituent parts, with (1,3) and (2,4) at coincidental locations. When the system is in a state of two well-separated molecules, $R$ measures the intermolecular separation. In this coordinate system, the exchange and parity symmetries of the system can also be readily imposed via boundary conditions for $\theta$ and $\phi$, as discussed in Appendix A. For further information on the hyperspherical representation, refer to Ref.~\cite{Rittenhouse:four-fermion}. 

With the hyperspherical coordinates, the Schr\"{o}dinger equation of the four-body system in 1D can be decomposed into a hyperangular component and a hyperradial component, giving the hyperspherical Schr\"{o}dinger equation for the reduced wavefunction $\psi(R,\theta,\phi) = R\;\Psi(R,\theta,\phi)$ (where $\Psi(R,\theta,\phi)$ is the wavefunction, and we set $\hbar = 1$ throughout this paper), 
\begin{equation}\label{eq6}
\Bigl[-\frac{1}{2\mu}\frac{\partial^2}{\partial R^2}+\hat{H}_{\text{ad}}(R,\theta,\phi)\Bigr]\psi(R,\theta,\phi) = E\psi(R,\theta,\phi)
\end{equation}
where $\hat{H}_{\text{ad}}$ is the adiabatic Hamiltonian,
\begin{equation}\label{eq7}
    \hat{H}_{\text{ad}}(R,\theta,\phi) = \frac{\hat{L}^2(\theta,\phi)}{2\mu R^2}+\hat{V}(R,\theta,\phi),
\end{equation}
$\hat{L}$ is the (hyper)angular momentum operator, and $\hat{V}$ is the interaction between the particles. $\hat{V}$ is obtained by summing up all the pairwise interactions $U(r)$ for each pair of molecules with separation $r$, i.e., $V(R,\theta,\phi) = \sum_{i>j}{U(r_{ij})}$.

It is useful to rewrite $\psi$ as
\begin{equation}\label{eq8}
    \psi(R,\theta,\phi) = \sum_{\nu}{F_{\nu}(R) \Phi_{\nu}(R;\theta,\phi)},
\end{equation}
where $\Phi_{\nu}(R;\theta,\phi)$ are eigenstates of the adiabatic Hamiltonian at $R$ labeled by $\nu$. The hyperspherical adiabatic potentials are defined as the corresponding eigenvalues $U_{\nu}(R)$ from
\begin{equation}\label{eq9}
    \hat{H}_{\text{ad}}(R,\theta,\phi)\Phi_{\nu}(R;\theta,\phi) = U_{\nu}(R)\Phi_{\nu}(R;\theta,\phi).
\end{equation}
Finally, $F_{\nu}(R)$ are the solutions to
\begin{align}\label{eq10}
&\Bigl[-\frac{1}{2\mu}\frac{d^2}{d R^2}+U_{\nu}(R)\Bigr]F_{\nu}(R)\\\nonumber
&-\frac{1}{2\mu}\sum_{\nu'}\Bigl[2P_{\nu\nu'}(R)\frac{d}{dR}+Q_{\nu\nu'}(R)\Bigr]F_{\nu'}(R)=E F_{\nu}(R)
\end{align}
where 
\begin{align}\label{eq11}
&P_{\nu\nu'}(R) = \langle\Phi_{\nu}(R,\theta,\phi)|\frac{\partial}{\partial R}| \Phi_{\nu'}(R,\theta,\phi)\rangle \nonumber\\
&= \int{d\phi\; d\theta\sin\theta\;\Phi_{\nu}(R,\theta,\phi)\frac{\partial}{\partial R}\Phi_{\nu'}(R,\theta,\phi)}
\end{align}and
\begin{align}\label{eq12}
&Q_{\nu\nu'}(R) = \langle\Phi_{\nu}(R,\theta,\phi)|\frac{\partial^2}{\partial R^2}| \Phi_{\nu'}(R,\theta,\phi)\rangle \nonumber\\
&= \int{d\phi\; d\theta\sin\theta\;\Phi_{\nu}(R,\theta,\phi)\frac{\partial^2}{\partial R^2}\Phi_{\nu'}(R,\theta,\phi)}.
\end{align}
The above reformulation of the Schr{\"o}dinger equation is exact, but in this paper we will follow previous literature on complex molecular collisions and concentrate much of our analysis on the adiabatic potential $U_\nu(R)$. When we do solve the Schr{\"o}dinger equation, we will restrict to the fully adiabatic (zero-coupling) approximation, setting $P_{\nu\nu'}=Q_{\nu\nu'}=0$.

To solve the hyperangular equation Eq.~\eqref{eq9}, B-splines are used as basis functions in both $\theta$ and $\phi$ coordinates. A detailed description is included in Appendix B. Similarly, the hyperradial equation~\eqref{eq10} is solved with B-splines in $R$. 

\begin{figure}
\centering
\includegraphics[width=0.48\textwidth]{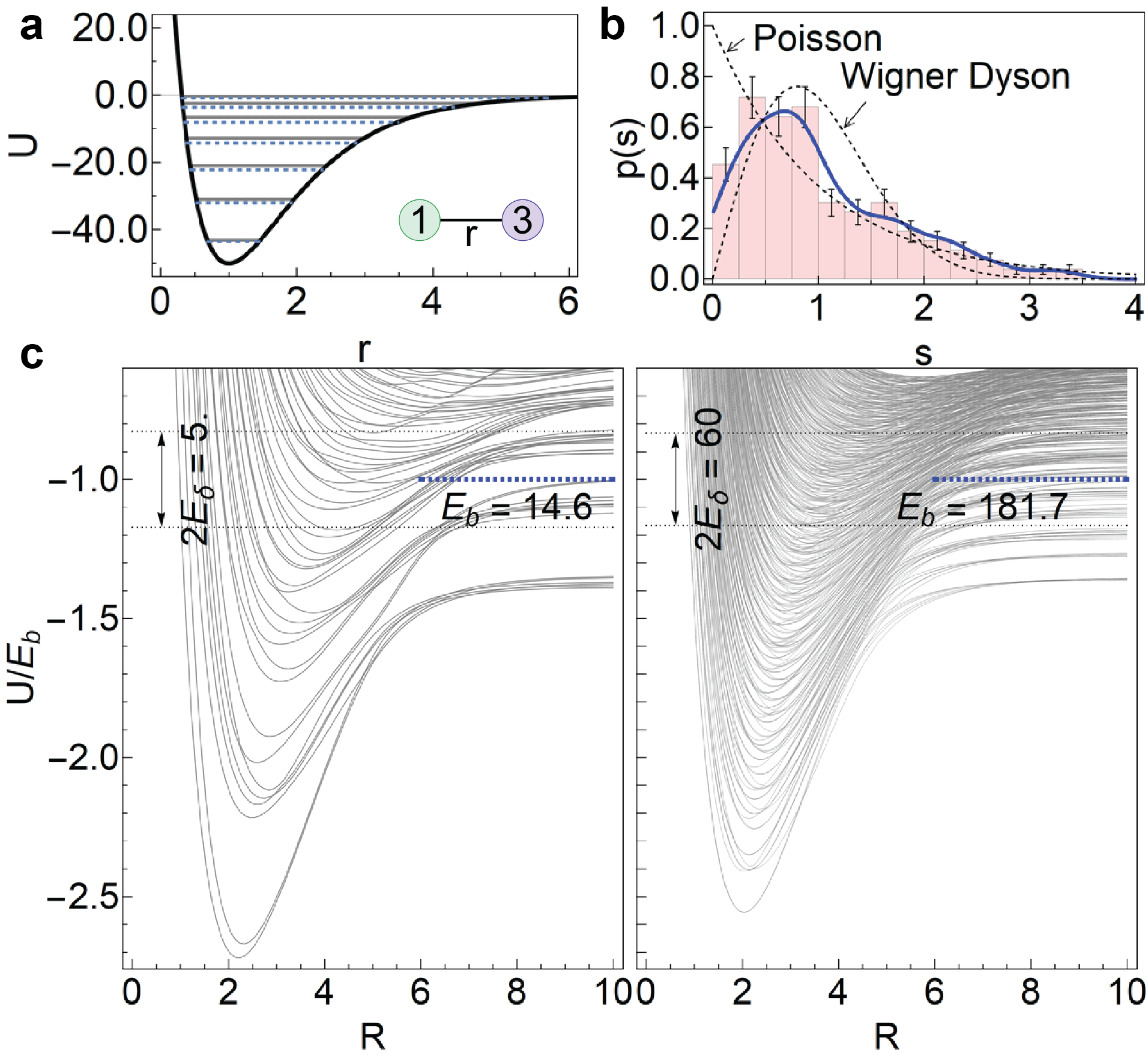}
\caption{\label{fig2} (color online) \textbf{(a)} Bound states of the two-atom interaction between atom 1 and 3 in a Morse potential, Eq.~\eqref{1second}, for $D = 50$, $a = 1/r_{0}$, and $r_{0} = 1$. Energy levels for $m_{3}/m_{1}=1$ (solid) and $m_{3}/m_{1}=1.3$ (dashed) are shown. \textbf{(b)} Probability distribution $p(s)$ of nearest neighbor spacing of adiabatic potential curves, scaled by the average spacing at hyperradius $R = 3.8$. Here $D = 100$, $r_{0} = 1$, and $m_3/m_1 = 1.3$. States are included from an energy range $-E_b\pm E_\delta$, where $E_b$ is the magnitude of the average lowest dimer-dimer energy at large separation, here and for all the statistical analyses in this paper. For this panel, $E_{\delta} = 30$. The dashed curves are the distributions for the Gaussian Orthogonal Ensemble and the Poisson distribution, respectively. The solid curve plots the kernel density estimator from the level spacing distribution data, using a Gaussian kernel with bandwidth 0.2. \textbf{(c)} The calculated hyperradial adiabatic potentials (solid lines) as a function of hyperradius $R$ at $r_{0} = 1$, $m_3/m_1 = 1.3$. $D = 10$ (left plot) and $D = 100$ (right plot) for the Morse potential. The right plot corresponds to the probability distribution plotted in \textbf{(b)}.}
\end{figure}
\section{Numerical results for adiabatic potentials}\label{sec:III}

Figure~\ref{fig2}(a) displays the two-atom bound states of Morse potential with $D= 50$ and $r_0 = 1$ for mass ratios $m_3/m_1 = 1$ and $m_3/m_1=1.3$. For each case, there are 14 bound states.

Figure~\ref{fig2}(c) shows the adiabatic potential curves $U_{\nu}(R)$ obtained by numerically solving the adiabatic equation at each hyperradius $R$ and imposing even parity boundary conditions (see Appendix A for more explanations on the parity symmetry of the system). The left and right plot in Fig.~\ref{fig2}(c) correspond to Morse potential with $D=10$ and $D=100$, respectively. The density of adiabatic potential curves increases as the number of bound states for the two-atom interaction increases. Since our interest is in collisions of ground-state molecules, the vertical axis is scaled by $E_b$, which is the absolute value of the average lowest dimer-dimer energy at large $R$. $E_b$ is the average of $|E_{12} + E_{34}|$ and $|E_{13} + E_{24}|$, the energies of two possible dimer-dimer configurations. The density of adiabatic potentials increases dramatically as the depth of the model potential increases, predicting a high density of four-body bound states. This will be discussed in detail later.

\begin{figure}
\centering
\includegraphics[width=0.48\textwidth]{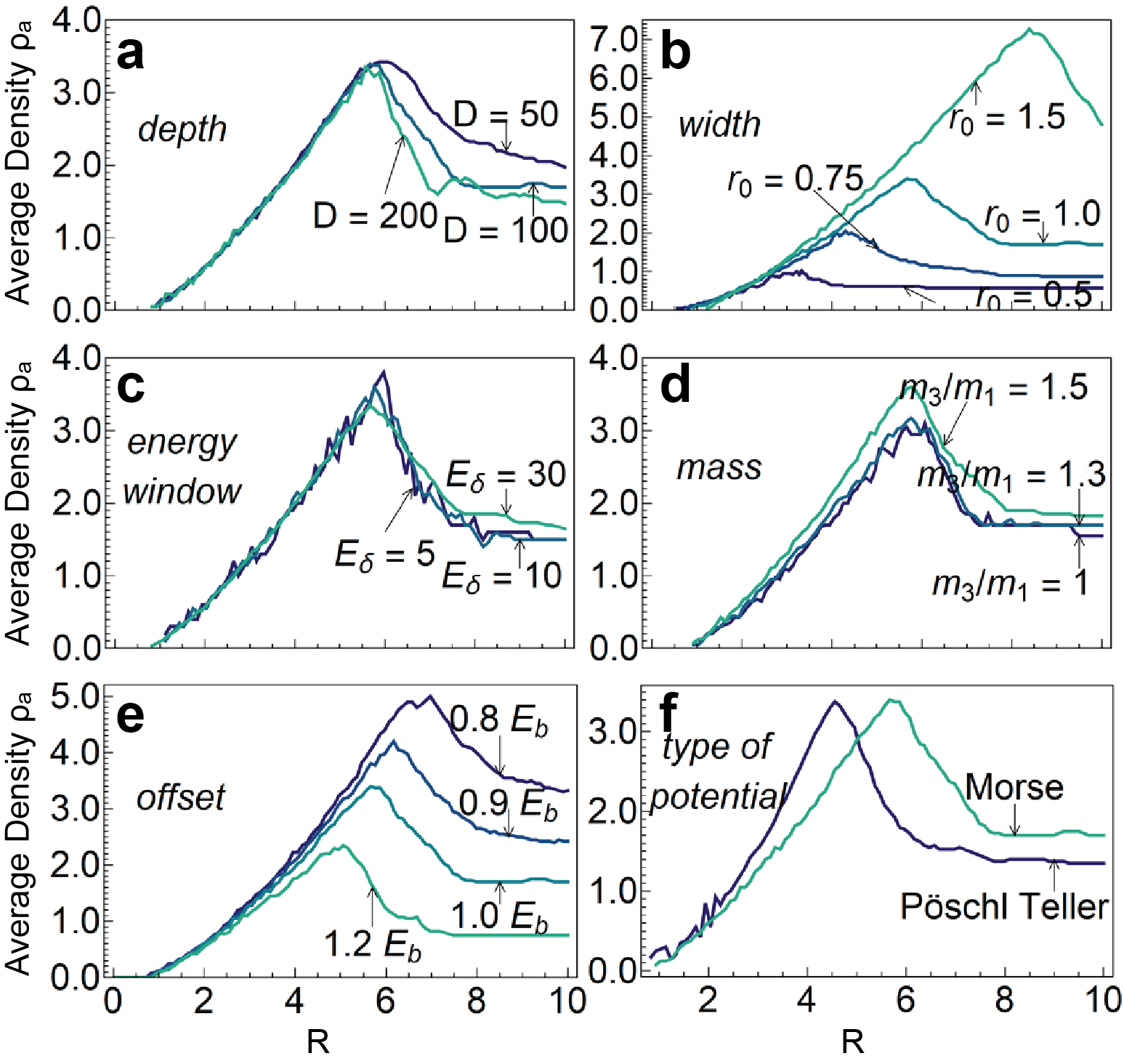}
\caption{\label{fig3}(color online) \textbf{(a)}-\textbf{(f)} illustrate the level density of adiabatic potential curves $\rho_a$ as a function of hyperradius $R$ for different two-atom interactions. Each plot examines the change in $\rho_a$ by varying one parameter of the two-atom interaction. The default non-varying parameters for \textbf{(a)}-\textbf{(f)} are set to $D = 100$, $r_0 =1.0$, $m_3/m_1 = 1.3$. The energy range of the included curves is $-E_b\pm E_{\delta}$, where $E_b = (E_{12} + E_{34} + E_{12} + E_{34})/2 = 181.7 $, and $E_{\delta}=30$. Morse potential is used in \textbf{(a)}-\textbf{(e)}; $r_0 =2$ for the P\"{o}schl-Teller potential curve in \textbf{(f)}. \textbf{(d)} $m_3/m_1 = 1$. Offset in \textbf{(e)} is defined as the center of the energy range.}
\end{figure}

The statistical distribution of adiabatic potentials $U_v(R)$ as a function of $R$ is particularly important for understanding quantum chaos~\cite{Chang:adiabatic_potential, Capecchi:potential_surfaces, Daily2015}, and one of the most informative and basic measures of the adiabatic potential statistics is shown in Fig.~\ref{fig2}(b). Fig.~\ref{fig2}(b) shows the probability distribution of nearest-neighbor spacing of the adiabatic potential curves at $R = 3.8$, for the $D=100$ case shown in Fig.~\ref{fig2}(c). The horizontal axis $s$ is scaled by the average level spacing. At the hyperradius shown, the level distribution is close to a chaotic distribution, with significant level repulsion readily apparent. 144 curves in an energy range $-E_b \pm E_\delta$ are included in the statistics, where $E_b= 181.7$ and $E_\delta=30$, so that the density of states is approximately uniform over the energy window. The solid curve is the kernel density estimation of the probability distribution given by $p(s) = 1/(\sqrt{2\pi}nh) \sum_{i=1}^{n} e^{-[(s-s_{i})/h]^2/2}$ where $s_i$ are data values and $h = 0.2$, an alternative useful to the histogram to estimate $p(s)$. The Wigner-Dyson distribution corresponding to a Gaussian Orthogonal Ensemble and the Poisson distribution, which are predicted to describe chaotic and integrable systems, respectively, are drawn in dashed curves for comparison. 

Figure~\ref{fig3} compares the level densities $\rho_a$ of the adiabatic potential curves as a function of hyperradius for different two-atom interactions, and demonstrates a universal trend of $\rho_a$ first increasing as $R$ increases, attaining a maximum, and then decreasing to a constant value at long range. The Morse potential was used for panels (a)-(e), where it is revealed that $\rho_a$ is independent of $D$ and $r_0$ for small $R$. This occurs because the kinetic energy dominates in this region, and therefore $\rho_a$ is independent of the two-atom interactions. As $R$ increases, $\rho_a$ first increases as a result of the decreasing splitting between eigenvalues of the kinetic energy ($\propto 1/R^2$), and then decreases and converges to a constant value as the spectrum converges to that of two independent molecules. The peak appears at around $R=6$ for Morse potential with $r_0 = 1$. Fig.~\ref{fig3}(a) and Fig.~\ref{fig3}(b) shows that $\rho_a$ decreases (increases) as the depth(width) of the potential increases. For reference, the number of the two-atom bound states for $D = 100$ and $D = 200$ are 20 and 28, respectively. Fig.~\ref{fig3}(c) demonstrates that $\rho_a$ is independent of the width of energy range $E_\delta$ over which the density is calculated for a range of widths from $E_\delta = 5$ to $30$. In Fig.~\ref{fig3}(d), no significant shift of $\rho_a$ is observed with a change in mass difference. Fig.~\ref{fig3}(e) shows that $\rho_a$ increases as the center of the energy window used to calculate $\rho$ shifts from low to high ($-1.2 E_b$ to $-0.8E_b$). This suggests the increase of the complexity of the adiabatic potential curves as the system shifts to higher energies. Fig.~\ref{fig3}(f) shows the similarity of these features for the Morse and P{\"o}schl-Teller potential.

\begin{figure}
\centering
\includegraphics[width=0.48\textwidth]{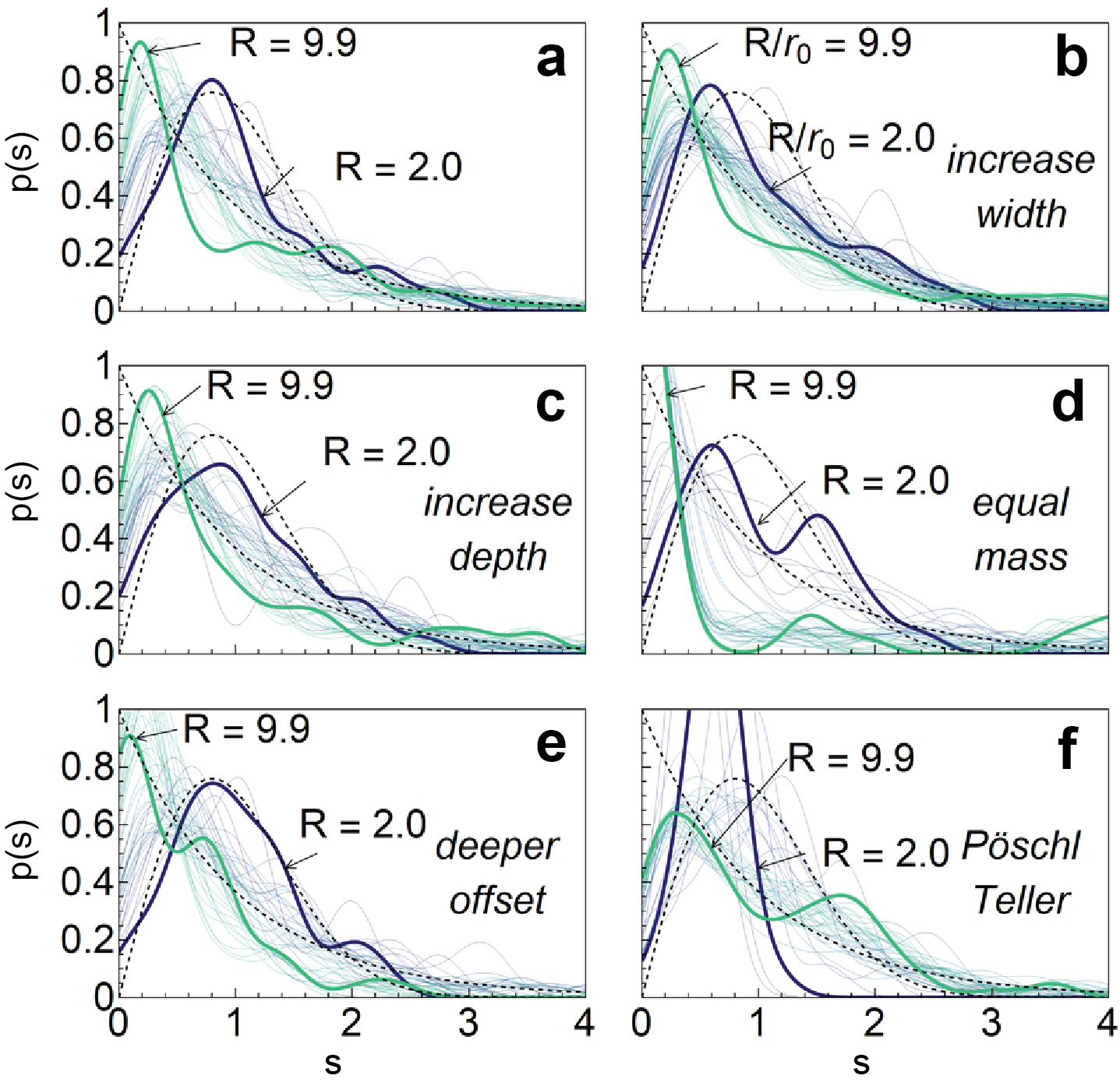}
\caption{\label{fig4} (color online) Probability distribution $p(s)$ of nearest neighbor spacing of adiabatic potential curves, scaled by the average spacing, at different hyperradii $R$. (Kernel density estimator obtained from data with a normal kernel of bandwidth 0.2.) The semi-transparent curves are for $R = 1.6$ to $R = 10$ in steps of $\Delta R = 0.2$; the bold curves plot $p(s)$ at $R=9.9$ and $R=2.0$. The dashed curves are the Poisson and GOE distributions in comparison. \textbf{(a)} The result for the Morse potential with $D=100$, $r_{0} = 1$, and $m_3/m_1=1.3$. The energy range of included curves is taken to be $- E_b\pm E_{\delta}$, and $E_{\delta} = 30$. \textbf{(b)-(e)} are various cases where one parameter (indicated in each plot) of the two-atom interaction in \textbf{(a)} is modified. \textbf{(b)} The width of the potential characterized by $r_0$ is set to $r_0$ = 2. \textbf{(c)} $D = 200$. \textbf{(d)} $m_3/m_1 = 1$. \textbf{(e)} The center of the included curves is shifted from $E_b$ to $E'_b = 1.2 E_b = 217.6$. \textbf{(f)} Using the P\"{o}schl-Teller potential with the same $D$, $m_3/m_1$, and $E_\delta$ as panel \textbf{(a)}. $r_{0} = 2$ is used to keep the numbers of bound states of the two-atom interactions the same for \textbf{(f)} and \textbf{(a)}.
}
\end{figure}

\begin{figure}
\centering
\includegraphics[width=0.48\textwidth]{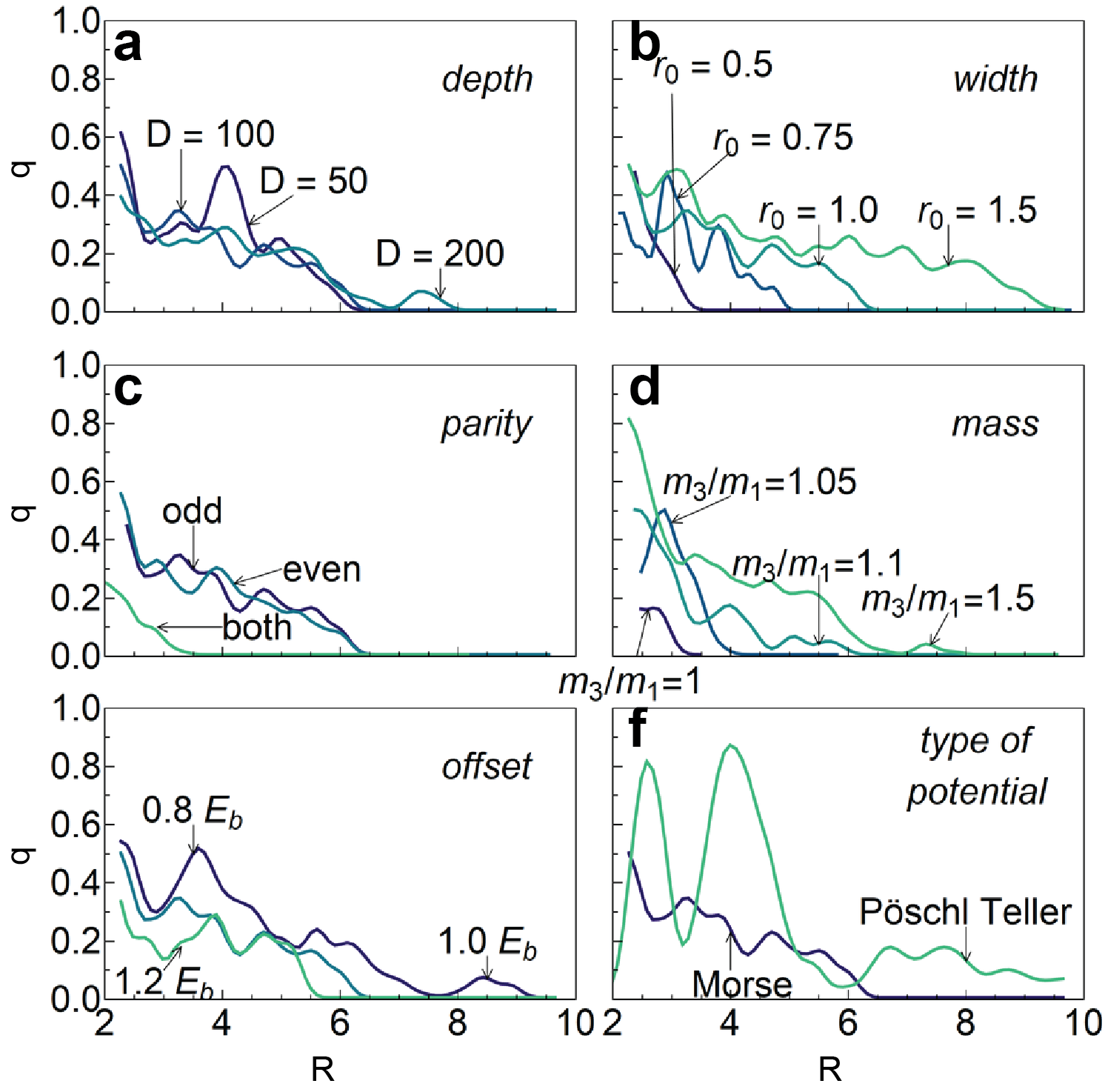}
\caption{\label{fig5} (color online) Brody parameter $q$ as a function of hyperradius $R$ for different two-atom interactions. Each plot examines the change in $q$ by varying one parameter of the two-atom interaction. The default parameters are the same from those in Fig.~\ref{fig3}. Note that the horizontal axis orign starts from $R = 2$. Results for different parities are shown in \textbf{(c)}. The plots are smoothed via moving average with weights from a Gaussian distribution.}
\end{figure}

Figure~\ref{fig4} reveals that the nearest-neighbor spacing distribution of adiabatic potentials evolves from the Wigner-Dyson distribution at $R=3.8$ towards a Poisson distribution as $R$ increases. For each panel in Fig.~\ref{fig4}, two curves are highlighted by thick lines to represent cases of small and large $R$, and the other lighter curves show other values of $R$ from $1.6$ to $10$. Each panel shows a different choice of model potential parameters or masses corresponding to the panels in Fig.~\ref{fig3}. The chaotic to non-chaotic crossover is clear in panels (a-c) and (e), though with some deviations to be examined later. In panel (d), a similar trend is visible, with level repulsion and a somewhat Wigner-Dyson-like distribution emerge at small $R$, but at large $R$ the distribution deviates dramatically from Poisson, showing a bunching of levels. Panel (f) is the only case where the level repulsion at small $R$ is questionable, showing only a modestly stronger repulsion than for large $R$. We will examine the origins of the features in panels (d) and (f) shortly. (Notice that in these plots, the small dips near $s=0$ that persist to large hyperradii are artifacts of the Gaussian smooth kernel distribution fitting. A more quantitative measure of the chaoticity of the statistical distribution will be presented later.)

A few other features emerge from close examination of Fig.~\ref{fig4}(a-f). In Fig.~\ref{fig4}(d), where $m_3/m_1=1$, a spike near $s=0$ arises for large $R$. This is the result of near-degeneracies that occur for large $R$ when $m_3/m_1=1$. In Fig.~\ref{fig4}(f), which shows a P{\"o}schl-Teller potential with $r_0=2$, chosen to have the same number of two-atom bound states as panel (a), there is still a visible level repulsion at small $R$ relative to large $R$, but the shape of the curves is always less well described by both the Wigner-Dyson and Poisson distributions. The difference in the short-range behavior for the P\"{o}schl-Teller case stems from the formation of bands of the adiabatic potential curves at short range. Since at small $R$, the value of P\"{o}schl-Teller potential does not vary rapidly, the potential energy surface is nearly constant in the hyperangular coordinates $\theta$ and $\phi$. Hence, the adiabatic potentials separate into bands, where each band is composed of potentials with the same kinetic energy eigenvalue, and the bandwidth -- the separation among these potential curves -- results from the small differences in their potential energy.

Figure~\ref{fig5} shows a more detailed evaluation of the evolution of chaoticity observed in the level spacing statistics as a function of $R$, by plotting the Brody parameter $q$~\cite{Brody:1973LettNuovoCimento, Brody:1981rmp} as a function of the hyperradius~\footnote{In Figure~\ref{fig5}, because the potential curves are selected from a specific energy range, $q$ has rapid oscillations as $R$ changes, as levels move in and out of the selected energy window. A moving average is applied to smooth the curves to determine their general feature. The average $(R_i,q_i)$ is given by $R_i = \sum_{n=i}^{n=i+9} f_n R_n$ and $q_i = \sum_{n=i}^{n=i+9} f_n q_n$, where $R_n$ and $q_n$ are the hyperradii and Brody parameter evaluated at grid points indexed by $n$, with adjacent $R_n$ separated by $\Delta R = 0.1$. $f_n$ are weights given by a normalized Gaussian with width 1.58.}. The Brody distribution is a statistical distribution characterized by $q\in [0,1]$ that can be expressed as
\begin{equation}\label{eq13}
    P_{\mathrm{Brody}}(s) =  {\Gamma\Big(\frac{2 + q}{1 + q}\Big)}^{q + 1} (1 + q)\; s^q \exp(-b s^{q + 1}).
\end{equation}It reduces to the standard Poisson distribution at $q = 0$ and to Wigner-Dyson at $q=1$. The $q$ plotted in Fig~\ref{fig5} is determined by a goodness-of-fit hypothesis test to Eq.~\eqref{eq13}.

Again, all cases shown in Fig~\ref{fig5} are broadly similar. Take Fig.~\ref{fig5}(a) as an example: $q$ smoothly decreases as $R$ increases, at some point falling to zero. The maximum $q$ occurs for small $R$, peaking around $q\sim 0.5$ for $D=100$. Fig.~\ref{fig5}(a) also shows no significant change in $q$ by changing the two-atom interaction strength over a range from $D = 50$ to $200$. Fig.~\ref{fig5}(b) conveys a similar message that $q$ is unaffected by a change in the width of two-atom potential from $r_0 = 0.5$ to $1.5$, after $R$ is rescaled by width $r_0$. Fig.~\ref{fig5}(c) shows that $q$ is roughly independent of parity, and that when adiabatic curves of both even and odd parities are included, they become doubly degenerate for $R \gtrsim4$, leading to a zero effective $q$. At small R, the value of q is approximately cut in half when both even and odd parity are included because those sets of spectral lines are uncorrelated with each other, and thus the characteristic level repulsion present in each set separately is not present in the combined set. Fig.~\ref{fig5}(d) exhibits an increase in $q$ as the difference in the mass of distinguishable fermions increases. Fig.~\ref{fig5}(e) shows that there is a small increase of $q$ as the system is examined at a higher energy. Fig.~\ref{fig5}(f) shows that the trends in $q$ for the P{\"o}schl-Teller potential are roughly similar, but $q$ is larger, and with very large oscillations. However the P{\"o}schl-Teller results are poorly behaved and caution must be exercised in interpreting the results.

\begin{figure}
\centering
\includegraphics[width=0.48\textwidth]{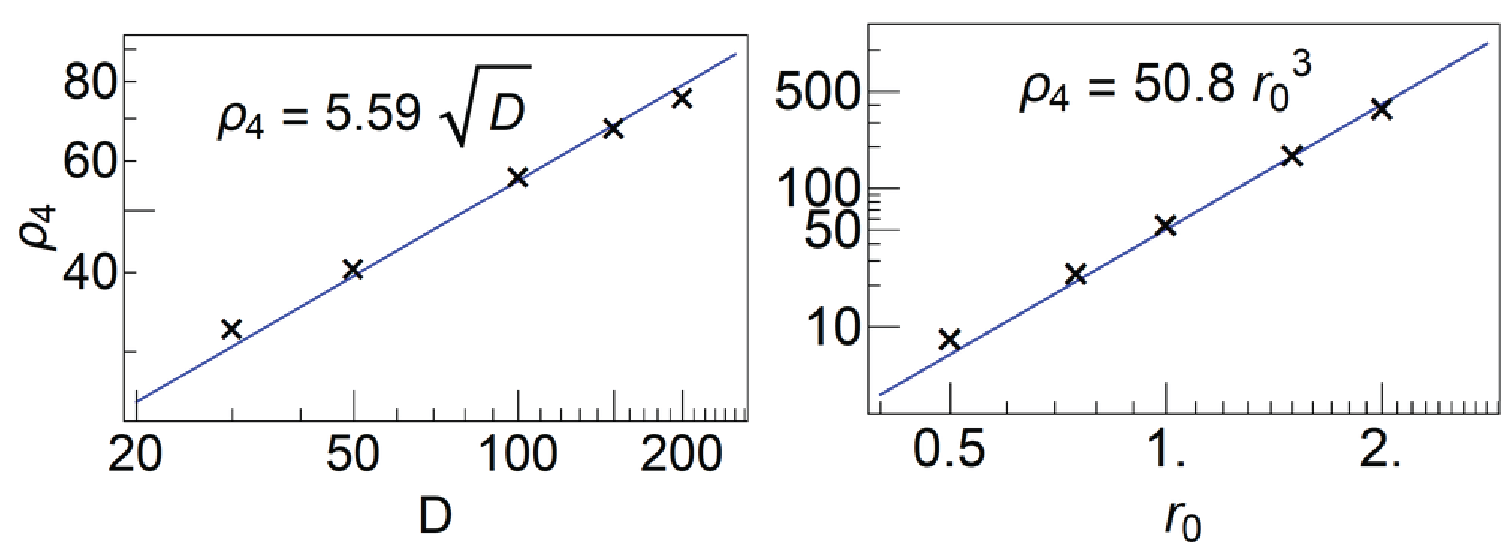}
\caption{\label{fig6} (color online) Estimated four-body bound states density $\rho_4$ as a function of \textbf{(a)} potential depth $D$ and \textbf{(b)} potential width $r_0$ for the Morse potential. Width $r_0 = 1$ is fixed \textbf{(a)}, and depth $D=100$ is fixed in \textbf{(b)}. The energy range of the four-body bound states are selected to be $E_{\Delta} = 3$, around which the fit converges. The points are fitted to show the linear and cubic relations.}
\end{figure}

\section{Four-body density of bound states: numerical results and analytic estimates}\label{sec:IV}
As mentioned in the introduction, an accurate, easy-to-use estimate of $\rho_4$ at the collision energy would be a powerful tool, for example, to determine the lifetime of cold molecule gases. Reference~\cite{Wall:2017pra2} derived an analytic formula, involving only simple power law dependences on the potential parameters, relating $\rho_4$ to the parameters of the atom-atom interaction for three-dimensional diatomic molecule collisions. A very recent paper by Groenenboom \textit{et al.}~\cite{Groenenboom:2019} derived a similar formula, properly incorporating the angular momentum conservation laws appropriate for zero-field scattering and incorporating more features of the potential energy surface. However, these estimates rely on several uncontrolled approximations whose reliabilities are difficult to gauge.

Here we derive the analogous relations for the one-dimensional problem that we consider, and show by comparison with numerical results that these relations capture the power law scaling of $\rho_4$ on two-atom interaction parameters, such as $D$ and $r_0$, very accurately. This provides evidence that the analogous analytic formulas for realistic three-dimensional molecular scattering may accurately capture the scaling of $\rho_4$ with molecular parameters.

We derive an analytic estimate of $\rho_4$ at collision energy $E$ starting with the following relationship:
\begin{equation}\label{eq15}
\rho_4 \propto N_4\rho_a(E).
\end{equation}
where $\rho_a(E)$ is the density of adiabatic channels $U_{\nu}(R)$ with threshold energy $E$, and $N_4$ is the number of four-body bound states supported by a single channel. In other words, $N_{4}$ is the number of bound states of an effective interaction $U_{\nu}(R)$.

For this approximation to hold, it is important to assume that the adiabatic potentials are identical except for a shift in energy. Although this approximation is substantial, it is roughly true -- the width and depth of the adiabatic potentials do not vary rapidly from channel to channel within the energy window of interest. Since at large $R$, the adiabatic potentials must asymptote to pairs of two-atom molecular bound states, their density can be related to the number of two-atom molecular bound states for the two-atom interaction, which we denote as $N_2$.

To estimate $\rho_a(E)$, we note the number of adiabatic channels with threshold between $E$ and $E+\Delta E$. This is the number of ways that a pair of diatomic molecules can have energy from $E$ to $E+\Delta E$ (assuming $E>-D$),
\begin{align}\label{eq16}
\rho_a(E) \Delta E = &\int_{-D}^{E+\Delta E} d\epsilon_1 \;\rho(\epsilon_1) \nonumber\\
&\int_{-D}^{E+\Delta E-\epsilon_1} d\epsilon_2 \; \rho(\epsilon_2) \; \Theta(\epsilon_1+\epsilon_2-E),
\end{align} 
where $\rho(E)$ is the two-atom density of states, and $\Theta$ is the Heaviside step function. We make another substantial approximation that $\rho(E)$ is roughly a constant independent of $E$, so $\rho \sim N_2/D$. Performing the integration in Eq.~\eqref{eq16} yields 
\begin{equation}\label{eq17}
\rho_a(E)\Delta E\sim \Big(\frac{N_{2}}{D}\Big)^2(D+E)\Delta E
\end{equation}

With the assumption that $U_{\nu}(R)$ has roughly the same shape as the two-atom interaction $U(r)$ (i.e. the shape may be different in details, but has a similar width and depth), we can approximate $N_4 = c_2 N_2$, where $c_2$ is an $O(1)$ constant roughly independent of molecule parameters (still another approximation). Combining with Eq.~\eqref{eq15} and Eq.~\eqref{eq17}, and taking the $E\rightarrow0$ limit for low energy collisions, we have
\begin{equation}\label{eq19}
\rho_4 \propto \frac{N_{2}^3}{D}.
\end{equation}

The final step is to obtain $N_2$'s dependence on the two-atom potential. By Levinson's theorem, $N_2$ of a given parity for angular momentum $l=0$ is the difference in the phase of scattering $\delta$ at zero and infinity energy
\begin{equation}\label{eq20}
    N_2 = \frac{1}{\pi}[\delta(E=0)-\delta(E=\infty)]
\end{equation}
with a reduced mass $\mu_2=1/2$.
Using the WKB approximation,
\begin{equation}\label{eq21}
    \delta(E) = \int_{r_t}^{\infty}dr\sqrt{E-U(r)} - \int_{0}^{\infty}dr\sqrt{E}
\end{equation} with $r_t$ the classical turning point. With $\delta(E=\infty)=0$,
\begin{equation}\label{eq22}
    N_{2} = \frac{1}{\pi}\int_{r_t}^{\infty}dr\sqrt{-U(r)} = \sqrt{D} r_0,
\end{equation}in the last step evaluating the integral explicitly for the Morse potential.

Finally, substituting Eq.~\eqref{eq22} into Eq.~\eqref{eq19}, we obtain a scaling relation of $\rho_4$ with the depth and width of the two-atom interaction
\begin{equation}\label{eq23}
\rho_4 \propto r^3_{0} \sqrt{D}.
\end{equation}

Despite these crude approximations, Figure~\ref{fig6} shows that the scaling predicted by Eq.~\eqref{eq23} captures the numerically determined $\rho_4$ remarkably well. Fig.~\ref{fig6} plots the predicted power laws of $\rho_4$ as a function of two-atom interaction depth $D$ and width $r_0$ with an overall multiplicative fitting parameter, showing good agreement with the numerical results.

\begin{figure}
\centering
\includegraphics[ width=0.3\textwidth]{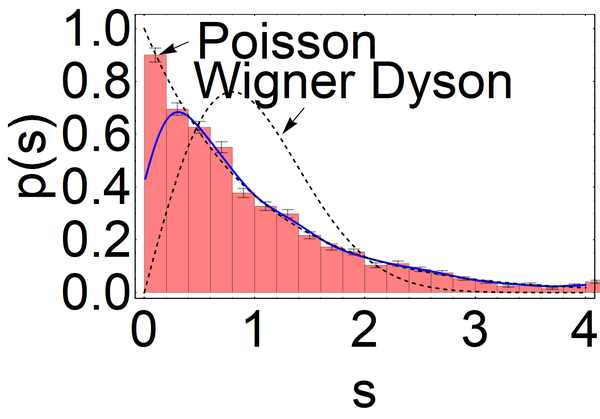}
\caption{\label{fig7} (color online) Probability density distribution $p(s)$ of nearest neighbor spacing of four-body bound states in the zero-coupling approximation for the Morse potential with $D = 100$, $r_{0} = 1$, $m_3/m_1 = 1.3$, with energy range $-E_b\pm E_{\delta}=-181.7\pm30$. The dashed curves correspond to Wigner-Dyson and Poisson distributions.}
\end{figure}
Finally, although we might expect that the chaotic (Wigner-Dyson) distribution of adiabatic channels would lead to a chaotic distribution of four-body bound states, Figure~\ref{fig7} shows that this is not the case -- at least within the approximation of zero nonadiabatic couplings. Rather, the nearest-neighbor spacing distribution of the approximated four-body bound states shows a nearly Poissonian distribution of levels, with no apparent signs of chaos. This is surprising given the clear signatures of chaos in the adiabatic potentials over a significant range of $R$. One possible explanation is that the relative spacing of each adiabatic potential curve is large compared to the level spacing of the four-body bound states, so the chaotic signature carried by the hyperangular component is hidden in the bound state distribution. It is likely that including channel couplings $P_{\nu\nu'}$ and $Q_{\nu\nu'}$ will lead to a chaotic distribution of the four-body eigenstates, which would indicate that chaos in the four-body eigenstates is not directly inherited from the adiabatic potentials, but through a mechanism that relies on the nonadiabatic couplings.

 \section{Conclusions} \label{sec:conclusions}
We have introduced a toy model to understand molecular collisions by considering a 1D system of two identical molecules with identical interactions between each pair of molecules. This allows us to explore a molecular system which has proliferating scattering resonances in a numerically exact fashion. We demonstrate that the hyperspherical adiabatic potentials transition from chaotic (Wigner-Dyson) to non-chaotic (Poisson) with increasing intermolecular separation for certain model potentials, despite the simplicity of the potentials, the reduced dimensionality, and even though the number of two-atom bound states is only $\sim 20$.

We further studied the dependence of the four-atom system's properties on the two-atom interaction by exploring the parameters space of the model potential. We show that the system at short range becomes more Wigner-Dyson-like as the mass ratio of distinguishable particles $m_3/m_1$ increases. For both the case of $m_3/m_1 = 1$ and the case including both inversion parities, the short-range chaos is suppressed by the extra degeneracy in the system. It is found that changing the depth $D$ or width $r_0$ of the two-atom interaction within the order of magnitude does not affect the chaoticity of collisions measured by the Brody parameter. The Brody parameter is found to be greater at higher collision energy. 

We also calculated and analyzed the four-body density of states $\rho_4$ in the uncoupled approximation. Two remarkable results emerge from this analysis of the four-body bound states. First, we derived a scaling relation $\rho_4 \propto \sqrt{D} r_0^3$ under various (uncontrolled) approximations, analogous to the arguments of Ref.~\cite{Wall:2017pra2} that relates the density of four-body bound states to the parmeters characterizing the two-atom interaction. Despite the uncontrolled nature of the approximations leading to the simple scaling relations for $\rho_4$, the scaling quantitatively agrees with the numerical calculations. Second, the nearest-neighbor level spacing distribution of the four-body bound states shows no signs of chaos despite the clear chaos in the hyperangular coordinates that manifests in the adiabatic potential curves. We anticipate that including the nonadiabatic couplings in future work will lead to level repulsion between states supported by different adiabatic potential curves, revealing a signature of chaos in the full four-body spectrum.

It will be illuminating to explore realistic collisions in three dimensions. Nevertheless, there are several implications of this study for real molecule scattering. Simple (power law) relations between the four-body bound state density and two-atom interaction are found and can be explained by a simple theory. We can speculate based on this evidence that the analogous power laws derived in Ref.~\cite{Wall:2017pra2} may accurately predict $\rho_4$ for realistic molecule scattering in 3D. If so, this can provide a potent guide for understanding and guiding molecule experiments. Another sage for 3D systems derived from this study is a caution that contrary to expectation, chaos in the distribution in the adiabatic channels does not necessarily translate into the four-body bound state distribution.
 
 \acknowledgements 
This material is based upon work supported with funds
from the Welch Foundation, grant no.~C-1872. K.~R.~A.~H
thanks the Aspen Center for Physics, which is supported
by the National Science Foundation grant PHY-1066293,
for its hospitality while part of this work was performed. 

\section*{Appendix A: Symmetry and Boundary Conditions}\label{sec:APPENDIXA}

The exchange symmetry of fermions and spatial inversion symmetry can be used to simplify the problem by reducing the range of angular coordinates of the Schr\"{o}dinger equation and they also determine the proper boundary conditions. We can define an operator $\hat{P}_{ij}$ that exchanges the particles $i$ and $j$. 
Acting $\hat{P}_{12}$ on each of the Jacobi coordinates $\rho_i$ gives
\begin{align}\label{A1}
    \hat{P}_{12}\rho_{1} &= -\rho_{1}\nonumber\\
   \hat{P}_{12}\rho_{2} &= \rho_{2}\tag{A1}\\
    \hat{P}_{12}\rho_{3} &= \rho_{3}\nonumber.
\end{align}
The hyperspherical coordinates $(R,\theta,\phi)$ transform as
\begin{align}\label{A2}
    \hat{P}_{12}R &= R\nonumber\\
   \hat{P}_{12}\theta &= \theta\tag{A2}\\
    \hat{P}_{12}\phi &= \phi-\pi\nonumber.
\end{align}
Since particles $1$ and $2$ are identical fermions, the wavefunction $\Phi(R,\theta, \phi)$ must be antisymmetric upon exchanging particles 1 and 2. Thus,
\begin{equation}\label{A3}
\hat{P}_{12}\Phi(R,\theta,\phi) = \Phi(R,\theta, \pi-\phi) = -\Phi(R,\theta,\phi). \tag{A3}
\end{equation}Similarly for $\hat{P}_{34}$,
\begin{equation}\label{A4}
\hat{P}_{34}\Phi(R,\theta,\phi) = \Phi(R,\theta,-\phi) = -\Phi(R,\theta,\phi). \tag{A4}
\end{equation}From the above two equations, the wavefunction defined on $\phi \in [-\pi, \pi)$ can be determined by knowing its value for $\phi\in [0, \pi/2] $, since based on Eq.~\eqref{A4}, the value of $\Phi(R,\theta,\phi)$ for $\phi \in [-\pi, 0]$ can be determined by that for $\phi \in [0,\pi]$, and based on Eq.~\eqref{A3}, $\Phi(R,\theta,\phi)$ for $\phi \in [\pi/2, \pi]$ can be determined by that for $\phi \in [0,\pi/2]$. From Eq.~\eqref{A3} and~\eqref{A4}, the boundary conditions on this reduced region are derived by letting $\phi \rightarrow 0$ or $\phi \rightarrow \pi/2$, resulting in $\Phi(R,\theta, \phi=0) = 0$ and $\Phi(R,\theta, \phi= \pi/2) = 0$. 

Spatial inversion symmetry $\hat{\Pi}$ also reduces the coordinate region. This symmetry yields $\rho_i \to -\rho_i$ for $i=1,2,3$, which corresponds to $(\theta,\phi)\to (\pi-\theta,\pi+\phi)$. The wavefunction must be invariant up to a phase under this symmetry, and $\hat{\Pi}^2 = 1$, so
\begin{equation}\label{A5}
\hat{\Pi} \Phi(R,\theta,\phi) = p\Phi(R,\theta,\phi)=\Phi(R,\pi-\theta,\pi+\phi),\tag{A5}
\end{equation}where $p = -1$ for odd parity, and $p = 1$ for even parity. By arguments similar to those above for the fermionic exchange, Eq.~\eqref{A5} implies that the range of $\theta$ can be reduced from $[0,\pi]$ to $[0, \pi/2]$. 

Lastly, an inversion of the system's geometry, $\rho_3 \rightarrow -\rho_3$, while fixing $\rho_1$ and $\rho_2$ (equivalent to applying $\hat{\Pi}\hat{P}_{12}\hat{P}_{34}$) should not alter the boundary conditions for the reduced region, i.e., 
\begin{equation}\label{A6}
\hat{\Pi}\hat{P}_{12}\hat{P}_{34} \Phi(R,\theta,\phi) = p\Phi(R,\theta,\phi)=\Phi(R,\pi-\theta,\phi),\tag{A6}
\end{equation}
Setting $\theta = \pi/2$, we arrive at another boundary condition: $\Phi(R,\theta,\phi)|_{\theta=\pi/2} = 0$ for odd parity and $\frac{\partial}{\partial\theta}\Phi(R,\theta,\phi)|_{\theta=\pi/2} = 0$ for even parity.

\section*{Appendix B: Solving the Schr\"{o}dinger Equation}\label{sec:APPENDIXB}
The hyperangular Schr\"{o}dinger equation 
\begin{align}\label{eqB1}
-&\frac{1}{2\mu R^2}\Big(\frac{1}{\sin{\theta}}\frac{\partial}{\partial\theta} \sin{\theta} \frac{\partial}{\partial\theta}+\frac{\partial^2}{\partial\phi^2} \Big)\Phi(R,\theta,\phi) \nonumber \\ &+ V(R,\theta,\phi) \Phi(R,\theta,\phi) = U(R) \Phi(R,\theta,\phi)\tag{B1}
\end{align}
is solved with B-spline bases~\cite{Liu:bsplines} for $R$, $\theta$, and $\phi$, by writing
\begin{equation}\label{eqB2}
\Phi(R,\theta,\phi) = \sum_{n}\sum_{m} c_{n,m} u_n(\phi) v_m(\theta).\tag{B2}
\end{equation}
$u_n(\phi)$ and $v_m(\theta)$ are B-spline functions. First, we insert the B-spline expansion in Eq.~\eqref{eqB2} into Eq.~\eqref{eqB1} and integrate both side with $\int{d\Omega}$ over the unit sphere, yielding
\begin{align*}
\sum_{n}\sum_{m}c_{n,m}&\Bigg\{-\frac{1}{2\mu R^2}\Big[s_{\phi}(n',n) t_{\theta}(m',m)\\
&+s_{\theta}(m',m)t_{\phi}(n',n)\Big]\\
&+\int{d\Omega \; u_n'(\phi) v_m'(\theta)\;V(R,\theta,\phi)\;u_n(\phi) v_m(\theta)}\Bigg\}
\\= &\sum_{n}\sum_{m}c_{n,m} U(R) s_{\phi}(n',n) t_{\theta}(m',m),\tag{B3}\label{eqB3}
\end{align*}
where
\begin{gather}
s_{\phi}(n',n)=\int{d \phi \; u_{n'}(\phi)u_{n}(\phi)}\nonumber\\
s_{\theta}(m',m)=\int{d \theta \; \sin{\theta} \;v_{m'}(\theta)v_{m}(\theta)}\nonumber\\
t_{\phi}(n',n)=\int{d \phi \; u_{n'}(\phi)\frac{\partial^2}{\partial\phi^2} u_{n}(\phi)}\tag{B4}\label{eqB4}\\
t_{\theta}(m',m)=-\int{d \theta \; \sin{\theta}\frac{\partial v_{m'}(\theta)}{\partial \theta} \frac{\partial v_{m}(\theta)}{\partial\theta}}.\nonumber
\end{gather}
The expression of $t_{\theta}(m',m)$ is obtained from integration by parts. Values in Eq.~\eqref{eqB4} are numerically integrated.

The hyperradial equation
\begin{equation}\label{eqB5}
\Bigl[-\frac{1}{2\mu}\frac{d^2}{d R^2}+U_{\nu}(R)\Bigr]F_{\nu}(R)=E F_{\nu}(R)\nonumber\tag{B5}
\end{equation}
is similarly solved on a B-spline basis. For all $R$, $\theta$, and $\phi$, B-splines are of order 5, and all numerical integrals such as those in Eq.~\eqref{eqB4} are evaluated with a Gaussian quadrature with 10 nodes. We choose the numbers of B-splines in $\theta$ and $\phi$ directions such that the error of the adiabatic potential curves at the largest plotted hyperradius is $\leq 1\%$ of the average spacing; the number in each direction varies from 100 to 160 depending on the specific parameters of the two-atom interaction. We numerically calculate the eigenvalues and eigenvectors in Eq.~\eqref{eqB3} and Eq.~\eqref{eqB5} using the Arnoldi algorithm in ARPACK.

\end{document}